\documentclass[a4paper, 12pt, fleqn]{article}
\usepackage[hmargin={1in, 1in}, vmargin={1in, 1in}]{geometry}
\usepackage[numbers, sort&compress]{natbib} 
\usepackage{color}        
\usepackage{graphicx}
\usepackage{amssymb}
\usepackage{epstopdf}
\usepackage{caption}	
\usepackage{mdwlist}
\begin{document}


\title{\begin{flushleft}Experimental verification of Boltzmann equilibrium for negative ions in weakly collisional electronegative plasmas\end{flushleft}}
\author{Young-chul Ghim$^{1}$ and Noah Hershkowitz$^2$}
\date{\begin{flushleft}
\textit{$^1$Rudolf Peierls Centre for Theoretical Physics, University of Oxford, Oxford, OX1 3NP, United Kingdom\newline
           $^2$Department of Engineering Physics, University of Wisconsin--Madison, Madison, Wisconsin, 53706, USA}
        \end{flushleft}}                                          

\maketitle
\thispagestyle{headings} 

\noindent
{\bf Abstract.}  
Weakly collisional Ar--O$_2$ electronegative plasmas are investigated in a dc multidipole chamber.  An electronegative core and an electropositive halo are observed.  The density ratio of negative ions to electrons ($\alpha$) in the nondrifting bulk is found to be 0.43.  The profile of $\alpha$ is found using both the phase velocity of ion acoustic waves and the drift velocity of positive ions determined by laser induced fluorescence.  The experiment shows that negative ions are in Boltzmann equilibrium with a temperature of $0.06 \pm 0.02\:eV$.  Double layers are not found separating the electronegative core and the electropositive halo.
\newline\newline


\indent
Experiments have demonstrated that electronegative plasmas are separated into three regions \cite{vender_pre_1995, kimura_jpd_1998, berezhnoj_apl_2000, plihon_pop_2008}.  These three regions are an electronegative core, an electropositive halo, and an ion rich sheath region  where the core and the halo in some parameter regimes \cite{franklin_jpd_1992, lichtenberg_jap_1994, sheridan_psst_1999, kouznetsov_jap_1999} are predicted to be separated by a double layer.  Braithwaite and Allen \cite{braithwaite_jpd_1988}, and Franklin \cite{franklin_psst_2002} argued that the appropriate parameters are $T_e/T_- > 10$ and $\alpha_0 \equiv n_{-0}/n_{e0} > (T_e/T_-)^{1/2}$.  Here, $T_e$, $T_-$, $n_{e0}$, and $n_{-0}$ denote the temperature of electron and negative ions in $eV$, and the number density of electron and negative ions in the bulk plasma, respectively.  Berezhnoj {\it et al.} \cite{berezhnoj_apl_2000} determined negative ion density profiles using photodetachment and Langmuir probe data in oxygen capacitively coupled rf plasmas.  They remarked that negative ions were in Boltzmann equilibrium for low pressure in comparing numerically calculated fluxes of negative ions and positive ions.  Plihon {\it et al.} \cite{plihon_pop_2008} reported the existence of double layers in electronegative Ar--SF${_6}$ plasmas.  However, these double layers were associated with a cusp magnetic field.  The double layers with potential steps of $8\:V$ for negative ion temperatures assumed to be at room temperature ($\sim\:0.025\:eV$) were not observed when electropositive plasmas were studied in the same system.
\newline\indent
Several authors have argued that it is appropriate to describe negative ions in low pressure electronegative plasmas by Boltzmann equilibrium \cite{lichtenberg_psst_1997, franklin_pop_2000, bogdanov_tpl_2001}.  However, direct measurements have not been carried out to verify if negative ions are in Boltzmann equilibrium in electronegative plasmas.  Spatial profiles of negative ion concentration and plasma potential are difficult to determine accurately.   This letter is the first experimental report to verify that negative ions are in Boltzmann equilibrium and that the double layer does not exist in weakly collisional electronegative plasmas for low negativity ($\alpha_0 = 0.43$).  The negative ion temperature is measured in the electronegative plasmas.
\newline\indent
The dispersion relation of ion acoustic waves (IAW) in electronegative plasmas can be derived from the general electrostatic wave dispersion relation \cite{hershkowitz_psst_2009},
\begin{equation}\label{eq:iaw_disp}
1=\sum_j\frac{\omega^2_{pj}}{\left(\omega-v_{dj}k\right)^2-v^2_{tj}k^2},
\end{equation}
where $\omega$ is the frequency, $k$ is the wavenumber, $v_d$ is the drift velocity, and $v_t$ is the thermal velocity.  The subscript $j$ denotes the $j$th species.  $\omega_p$ is the plasma frequency, and the sum is over all species.  In electronegative plasmas with non-drifting ions, D'Angelo {\it et al.} \cite{dangelo_pf_1966} predicted that there exist two modes (fast mode and slow mode), and two modes were observed by Sato {\it et al.} \cite{sato_pop_1994}.  If positive ions drift in electronegative plasmas, then each mode has two different phase velocities corresponding to the IAW traveling in the direction (parallel mode) and in the opposite direction (antiparallel mode) of the ion drift.
\newline\indent
The fast mode IAW dispersion relation can be derived from Eq. (\ref{eq:iaw_disp}) by assuming $\omega/k \gg v_{t+}$, $\omega/k \gg v_{t-}$, $\omega/k \ll v_{te}$, and $k\lambda_{De} \ll 1$ where $\lambda_D$ is the Debye length.  Here, the subscripts $+$, $-$, and $e$ denote positive ions, negative ions, and electrons, respectively.  Making these assumptions, the fast mode IAW phase velocities ($v_{phF}$) with nonzero $v_{d+}$ are
\begin{equation}\label{eq:fast_iaw}
v_{phF}^2=c^2_S\left[\left(1+\alpha\right)\frac{v^2_{phF}}{\left(v_{phF}-v_{d+}\right)^2}+\alpha\frac{m_+}{m_-} \right],
\end{equation}
where $c_S=\sqrt{T_e/m_+}$, and $\alpha=n_-/n_e$.
\newline\indent
The slow mode IAW dispersion relation can also be derived from Eq. (\ref{eq:iaw_disp}) by assuming $\omega/k \gg v_{t+}$, $\omega/k \ll v_{t-}$, $\omega/k \ll v_{te}$ $k\lambda_{De} \ll 1$, and $k\lambda_{D-} \ll 1$.  With the above assumptions and nonzero $v_{d+}$, the slow mode IAW phase velocities ($v_{phS}$) are
\begin{equation}\label{eq:slow_iaw}
v_{phS}=v_{d+}\pm c_S\sqrt{\frac{1+\alpha}{1+\alpha\frac{T_e}{T_-}}}.
\end{equation}
In RHS of Eq. (\ref{eq:slow_iaw}), the `$+$' sign corresponds to the parallel mode, and the `$-$' sign corresponds to the antiparallel mode of the slow mode IAW.
\newline\indent
The IAW dispersion relation in electropositive plasmas is derived from Eq. (\ref{eq:iaw_disp}) by assuming $\omega/k \gg v_{t+}$, $\omega/k \ll v_{te}$, and $k\lambda_{De} \ll 1$ with nonzero $v_{d+}$:
\begin{equation}\label{eq:iaw_pos}
v_{ph}=v_{d+}\pm c_S,
\end{equation}
where $v_{ph}$ is the phase velocity of the IAW in electropositive plasmas.  The parallel (+) mode and the antiparallel (-) mode exist due to drifting ions.
\newline\indent
In electropositive plasmas, the location of the sheath edge is found to be where the parallel mode of the $v_{ph}$ is twice $c_S$ \cite{oksuz_psst_2008}.  It is equivalent to argue that the antiparallel mode of the $v_{ph}$ is zero at the sheath edge.  The same argument has employed to locate the inner sheath edge of electronegative plasmas.  Braithwaite and Allen \cite{braithwaite_jpd_1988} argued that at the inner sheath edge, the parallel mode of $v_{phS}$ is twice $c_S\sqrt{\left(1+\alpha\right)/\left(1+\alpha T_e/T-\right)}$, whereas the antiparallel mode of $v_{phS}$ becomes zero.  This argument is valid if a double layer exists at the inner sheath edge.
\newline\indent
The experiment was carried in a cylindrical dc multidipole chamber $70\:cm$ in length and $60\:cm$ in diameter (interior dimensions).  A detailed description of the chamber can be found elsewhere \cite{hoskinson_psst_2006}.   A stainless steel plate diameter of $15\:cm$ is biased to $-30\:V$ in the center of the chamber to act as a boundary.  Electronegative plasmas were generated by primary electrons emitted from thoriated hot filaments biased at $-60\:V$ and the discharge current $0.8\:A$ with employed neutral gases of Ar at $0.10\:mTorr$ and O$_2$ at $0.30\:mTorr$.  A planar Langmuir probe (LP) with radius of $0.3\:cm$ was utilized to provide the electron number density and the effective electron temperature in the bulk plasmas.  In this experiment, the LP trace showed the existence of hot and cold electrons.  Thus, the effective electron temperature was used to find the Bohm speed, $c_S$ \cite{song_pre_1997}.  From the LP trace it is found that $n_e \sim 3.8\times 10^9\:cm^{-3}$ and $T_e \sim 0.69\:eV$ which results in $c_S \sim 1290 \pm 50\:m/s$ with $m_+$ being the mass of Ar.  Use of the mass of Ar for $m_+$ is justified in the following paragraph.
\newline\indent
A tunable diode laser was used to determine the profile of Ar$^+$ drift velocity ($v_{d+}$).  The laser of $668.614\:nm$ (in vacuum) excited the metastable Ar$^+$ ($3d^4F_{7/2}$) to $4p^4D_{5/2}$ level, which in turn emits fluorescence radiation with a wavelength of $442.72\:nm$ (in air).  A detailed description of the LIF system can be found elsewhere \cite{lee_jpd_2006}.  It is assumed that the metastable Ar$^+$ is in thermal equilibrium with the ground state Ar$^+$ \cite{goeckner_pfb_1991}.  By scanning the wavelength of the laser, the ion velocity distribution functions (ivdf) were obtained from the Doppler shift.  The values of $v_{d+}$ are found by taking the first moment of the ivdfs.  
\begin{figure}[!t]
\centering
\includegraphics[width=4.5in]{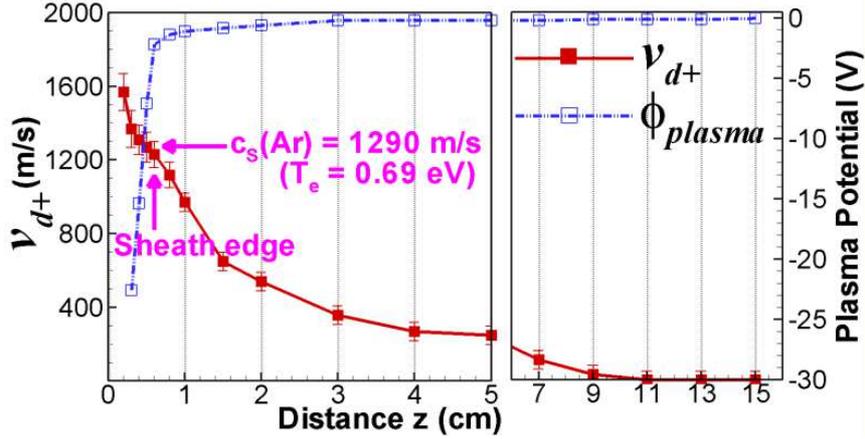}
\caption{\label{fig:vd_profile} (Color online) Profiles of $v_{d+}$ and $\phi_{plasma}$ from $z=15.0\:cm$ to $0.2\:cm$.}
\end{figure}
Figure \ref{fig:vd_profile} shows the profiles of the $v_{d+}$ and the plasma potential ($\phi_{plasma}$) measured by an emissive probe using the inflection point method in the limit of zero emission \cite{smith_rsi_1979}.  The sheath edge was located at $z = 0.6\:cm$.  Here, $z$ is the distance from the boundary to the diagnosed point.  The measured $v_{d+}$ at $z = 0.6\:cm$ was $1230 \pm 70\:m/s$.  This velocity agrees well with the calculated $c_S$ with the assumption of $m_+$ equal to the mass of Ar.  If there were non-negligible O$^+$ or O$^{2+}$, then the measured $v_{d+}$ would have been faster than the calculated $c_S$ \cite{lee_apl_2007}. 
\newline\indent
The values of the phase velocity of the IAW were measured using CW (continuous wave) excitation at $100\:kHz$ with peak-to-peak voltage of $3.0\:V$ and offset voltage of $0.5\:V$.  A detailed description of the CW IAW setup can be found elsewhere \cite{oksuz_psst_2008}.  The measured phase velocity is denoted by $v_{phM}$.  The excitation was launched at the center of the chamber with the launching grid diameter of $10\:cm$.  The LP is located between the launcher and the boundary which were separated by $20\:cm$.  The LP was positively biased so that it collected electron saturation current and was moved from $z = 15.0\:cm$ to $0.5\:cm$.  Here, $z$ is, again, the distance from the boundary.  The signal collected by the LP was averaged by a boxcar averager with $5\:kHz$ triggering, $30\:nsec$ window width, and delay time varied from $0.0$ to $10.0\:\mu sec$ with the step size of $0.5\:\mu sec$.  
\begin{figure}[!t]
\centering
\includegraphics[width=3.5in]{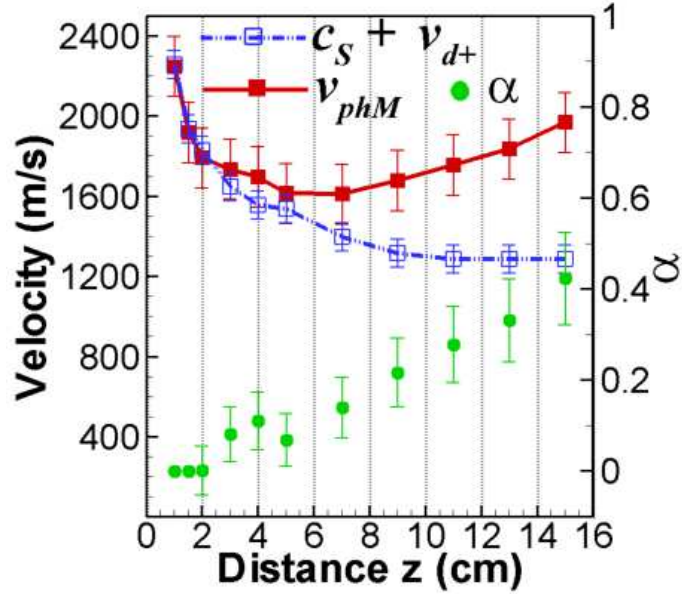}
\caption{\label{fig:vphm_profile} (Color online) The profiles of $v_{phM}$ and $c_S+v_{d+}=v_{ph}$.  The values of $\alpha$ are calculated using Eq. (\ref{eq:fast_iaw}) with the measured $v_{phM}$, $v_{d+}$ and $c_S$.}
\end{figure}
Figure \ref{fig:vphm_profile} shows the profile of $v_{phM}$ from $z = 15.0\:cm$ to $1.0\:cm$.
\newline\indent
The electronegative core and electropositive halo regions are identified by comparing $v_{phM}$ to Eq. (\ref{eq:iaw_pos}).  In the electronegative core, the values of $v_{phM}$ are not equal to $v_{ph}$ in Eq. (\ref{eq:iaw_pos}), whereas in the electropositive halo, these two velocities are equal.  Figure \ref{fig:vphm_profile} shows that these two velocities become the same at $z = 2.0\:cm$.  Thus, the electropositive halo starts at $z = 2.0\:cm$.  Consequently, from $z = 15.0$ to $2.0\:cm$, the measured phase velocity $v_{phM}$ corresponds to the fast mode of the IAW ($v_{phF}$) because this region is the electronegative core.  Using Eq. (\ref{eq:fast_iaw}) and having the measured values of $v_{phM}$ ($= v_{phF}$), $c_S$, and $v_{d+}$, $\alpha$ can be calculated.  Here, the dominant negative ions are assumed to be O$^-$ (i.e., $m_- = 16$) \cite{vender_pre_1995, berezhnoj_apl_2000}.  Figure \ref{fig:vphm_profile} shows the $\alpha$ profile and the existence of the electronegative core and the electropositive halo. 
\newline\indent
To verify the Boltzmann equilibrium for the negative ions, the Boltzmann relation is examined over the measured values of $\alpha$ and plasma potentials.  
\begin{figure}[!t]
\centering
\includegraphics[width=3.5in]{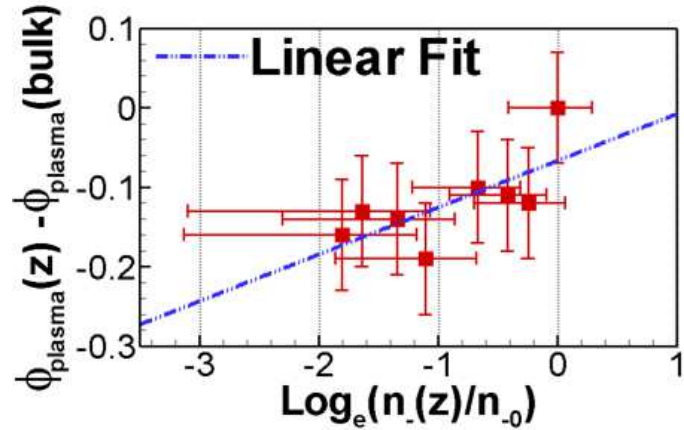}
\caption{\label{fig:boltzmann} (Color online) Negative ions follow the Boltzmann relation with $T_-=0.06\pm 0.02\:eV$.}
\end{figure}
Figure \ref{fig:boltzmann} shows that the negative ions agree with the Boltzmann relation with $T_- = 0.06 \pm 0.02\:eV$ within the experimental errors.  In addition, the plasma potential profile shows that there is approximately a $0.5 \pm 0.1\:V$ drop from $z = 15.0$ to $2.0\:cm$.  This potential drop is sufficient enough to confine the negative ions with the temperature of $0.06 \pm 0.02\:eV$ within the core.
\newline\indent
An electronegative core and an electropositive halo were detected.  Nevertheless, the plasma potential profile shows that there are no double layers.  The fact that the negative ion fraction is gradually decreasing (Fig. \ref{fig:vphm_profile}) rather than making an abrupt reduction suggests nonexistence of double layers.  The absence of double layers can also be verified by considering the positive ion drift velocity $v_{d+}$.  Eq. (\ref{eq:vd+}) is the theoretical prediction of the $v_{d+}$ at the inner sheath edge when there is a double layer.
\begin{equation}\label{eq:vd+}
v_{d+}=c_S\sqrt{\frac{1+\alpha}{1+\alpha\frac{T_e}{T_-}}}.
\end{equation}
Comparing the measured $v_{d+}$ with the RHS of Eq. (\ref{eq:vd+}) which is the $v_{phS}$ with nondrifting ions, 
\begin{figure}[!t]
\centering
\includegraphics[width=3.5in]{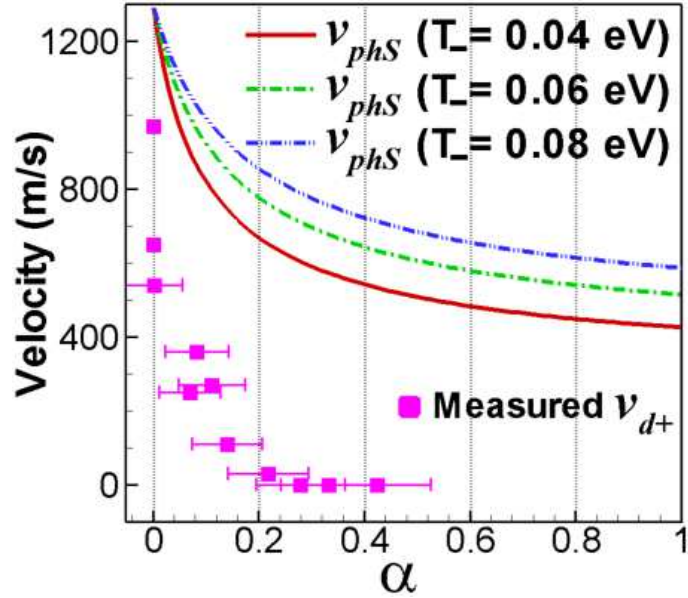}
\caption{\label{fig:vph_test} (Color online) RHS of Eq. (\ref{eq:vd+}) as a function of $\alpha$ with $T_e=0.69\:eV$, $T_-=0.04$, $0.06$ and $0.08\:eV$ (lines), and measured $v_{d+}$ as a function of $\alpha$ (square dots).}
\end{figure}
Fig. \ref{fig:vph_test} shows that $v_{d+}$ never reaches $c_S\sqrt{\left(1+\alpha\right)/\left(1+\alpha T_e/T_- \right)}$.  Thus, it can be concluded that the double layer for the investigated condition does not exist.  This is consistent with the theoretical prediction \cite{braithwaite_jpd_1988, franklin_psst_2002}.
\newline\indent
In conclusions, using the measured drift velocity of positive ions and the phase velocity of the IAW in electronegative plasmas, the profile of the negative ion fraction is found in a dc multidipole chamber.  The location where the electronegative core and electropositive halo are separated is found at $z = 2.0\:cm$, while the electropositive halo terminates at $z = 0.6\:cm$ followed by the sheath.  It is found that the negative ions follow the Boltzmann relation with a temperature of $0.06 \pm 0.02\:eV$ and that a double layer does not exist when negative ions are confined in the core of the plasmas.
\newline\newline\indent
We would like to thank Chi-shung Yip for his assistance of taking data.
This work was supported by U.S. DOE Grant No. DE-FG02-97ER54437.

\bibliographystyle{unsrt}
\setlength{\bibsep}{0.0pt}


\end{document}